\documentclass[prd,showpacs,aps,preprint]{revtex4}

\let\ssection=\section
\renewcommand{\section}{\setcounter{equation}{0}\ssection}

\begin{document}
\title{On the dynamics of classical particle with spin}

\author{Natalia Kudryashova}
\email{n.kudryashova@ucl.ac.uk}
\affiliation{Department of Mathematics,
University College London, Gower Street, London, WC1E 6BT, United Kingdom}
\author{Yuri N.\ Obukhov}
\email{obukhov@math.ucl.ac.uk}
\affiliation{Department of Mathematics and Institute of Origins, 
University College London, Gower Street, London, WC1E 6BT, United Kingdom}

\bigskip
\bigskip

\begin{abstract}
The complete explicitly covariant 4-dimensional description of the dynamics 
of a free classical particle with spin within the framework of the special 
relativity theory is presented. The key point of our approach is the
the introduction of the new vector field which enables to define the
analogues of the mean spin and position variables. The supplementary 
conditions are discussed and it is demonstrated that the Frenkel condition
unambiguously determines the dynamics of a spinning particle. 
\end{abstract}

\bigskip\bigskip
\bigskip
\maketitle

\section{Introduction}

The models of classical spinning matter go back to 1926 when, shortly after
Uhlenbeck and Goudsmit \cite{uhlen} have introduced the notion of intrinsic
angular momentum in quantum mechanics, J. Frenkel \cite{frenkel} developed
the theory of a classical electron with spin moving in electromagnetic 
field. Since that time, the dynamics of classical spinning matter (particles 
and continuous media) in special and general relativity was studied intensively.
For the overview see \cite{review}. 

Although the true understanding of spin as an essentially quantum property of 
matter is achieved only in quantum theory, the classical models nevertheless
turn out to be rich and consistent to an extent which justifies their 
applications for the analysis of the experiments with the high energy beams 
of polarised particles and with the polarised targets. That direction was 
brought to life by the work of Bargmann, Michel, and Telegdi \cite{bmt} who
established that the rigorous quantum theory practically exactly reproduces
the Frenkel model when the spin variable is defined as a mean value of
the quantum spin operator over a quasi-classical wave packet state. 

Considerable attention was paid to the study of dynamics of spinning particles 
in the gravitational field, see for example \cite{mash,moh,7r}, a historic
account and the list of the references can be found in \cite{dp}. Among other 
issues, the choice of the so-called supplementary condition was a subject of
a long discussion. The two main options are Frenkel's and Tulczyjew's 
conditions. Comparing these choices, one finds the statements in the literature
about the ``ambiguity'' allowed for particle's trajectory under the Frenkel 
condition \cite{tulczyjew,7r}. We will address this issue here.

In this paper we reconsider the dynamics of a classical particle with spin
within the framework of the special relativity theory. Although much of the
work was done previously, see, e.g., \cite{wey1,cor}, the problem was usually 
considered within the 3-dimensional framework (often using special reference
systems). Our aim is to describe the explicitly covariant 4-dimensional general
picture of a spinning particle dynamics. As a result, we will find certain new 
features, especially by making a careful comparison of Frenkel's and 
Tulczyjew's supplementary conditions. We demonstrate that the use of the
Frenkel condition yields a trajectory uniquely determined by the initial 
values of the momentum, spin and particle's position.

\section{Classical model of particle with spin}

We will confine our attention to a point particle with spin described by
the Frenkel-Weyssenhoff theory \cite{frenkel,wey1,cor}. In this classical 
model, a spinning particle is described as a physical point which position 
in the Minkowski spacetime is given by the four coordinates,
\begin{equation}
X^\mu\,=X^\mu(\tau),
\end{equation}
and which is characterised by the two gravitational ``charges'', the 
4-momentum vector and the spin tensor:
\begin{eqnarray}
P_{\alpha}&=&P_{\alpha}(X(\tau)),\\
S_{\alpha\beta}&=&S_{\alpha\beta}(X(\tau)).
\end{eqnarray}
The parameter $\tau$ can be chosen as an arbitrary Lorentz-invariant real 
variable. Without loosing generality, we will assign $\tau=0$ to the initial 
position of a particle. As usual, we also assume that $\tau$ is equal to the 
proper time along the world line of a particle, so that the 4-velocity,
\begin{equation}
U^\mu = {\frac {dX^\mu} {d\tau}},\label{velo}
\end{equation}
is normalised by the condition
\begin{equation}
U^\mu\,U_\mu=1.\label{norm}
\end{equation}

Finally, we assume the supplementary condition
\begin{equation}\label{cond}
S_{\alpha \beta } \left(\hat{\alpha }U^{\beta } +\hat{\beta }P^{\beta }\right)=0.
\end{equation}
with the constant parameters $\hat{\alpha}$ and $\hat{\beta}$. This general 
condition includes as limiting cases a so-called Frenkel \cite{frenkel} 
(sometimes also called Pirani \cite{pirani}) and Tulczyjew conditions 
\cite{tulczyjew}, with $\hat{\beta}=0$ and $\hat{\alpha}=0$, respectively. 
The Frenkel condition corresponds to the spin of pure ``magnetic" nature, and 
physically means that in the rest frame of a particle the spin has only three 
spatial components. Under the Tulczyjew's condition, the trajectory of an 
extended test body is determined by the position of the centre of mass of 
the body.
 
Dynamics of a free particle is governed by the standard conservation laws
of momentum and of the total angular momentum,
\begin{eqnarray}
{\frac d {d\tau}}P_{\alpha}&=&0,\label{momcon}\\
{\frac d {d\tau}}(X_\alpha\,P_\beta - X_\beta\,P_\alpha + 
S_{\alpha\beta})&=&0.\label{angcon}
\end{eqnarray}
Although this system looks rather simple, its complete integration is 
nontrivial. As a first step, denoting the initial values of the particle
position, momentum, and spin by ${\stackrel{(0)} X}{}^\mu,\,
{\stackrel{(0)}P}{}_\alpha,\,{\stackrel{(0)} S}{}_{\alpha\beta}$,
respectively, we find from the first integrals from 
(\ref{momcon})-(\ref{angcon}):
\begin{eqnarray}
P_\alpha&=&{\stackrel{(0)} P}{}_\alpha,\\
X_\alpha\,P_\beta - X_\beta\,P_\alpha + S_{\alpha\beta}&=&
J_{\alpha\beta},
\end{eqnarray}
with constant tensor
\begin{equation}
J_{\alpha\beta}:={\stackrel{(0)} X}_\alpha\,{\stackrel{(0)} P}_\beta 
- {\stackrel{(0)}X}_\beta\,{\stackrel{(0)} P}_\alpha + 
{\stackrel{(0)} S}_{\alpha\beta}.\label{Jab}
\end{equation}

For $(4+4+4+6)=18$ unknown variables $X^\mu,\,U^\mu,\,P_\alpha,\,S_{\alpha\beta}$,
we have exactly $(4+1+3+4+6)=18$ differential and algebraic equations, 
(\ref{velo})-(\ref{angcon}). Or, by substituting (\ref{velo}) into (\ref{norm})
and (\ref{cond}), we have a system of 14 differential-algebraic equations 
for the 14 variables $X^\mu,\,P_\alpha,\,S_{\alpha\beta}$. Accordingly, we expect
that all these dynamical variables are uniquely determined by their initial
values. 

Combining (\ref{angcon}) and (\ref{velo}), we find:
\begin{equation}
\dot{S}_{\alpha\beta}= - U_\alpha P_\beta + U_\beta P_\alpha.\label{angcon1}
\end{equation}
Hereafter the proper
time derivative will be denoted by the dot, ${\frac d {d\tau}}=(\dot{\ })$.
The system (\ref{momcon}) and (\ref{angcon1}) was generalised to the case
of the curved spacetime by Mathisson and Papapetrou \cite{mathpapa} (then
the curvature-dependent forces appear on the right-hand side). 
Contracting (\ref{angcon1}) with $U^\beta$, we get
\begin{equation}
P_\alpha = m_0 U_\alpha + U^\beta\dot{S}_{\alpha\beta},\label{P1}
\end{equation}
where we denote
\begin{equation}
m_0:=U^\alpha P_\alpha.
\end{equation}

It is easy to see that for the Frenkel and the Tulczyjew  conditions $m_0$ 
is constant along the world line [contract (\ref{P1}) with $\dot{U}^\alpha$ 
and use (\ref{norm}) and (\ref{cond}) to obtain $\dot{U}^\alpha P_\alpha=
\dot{m}_0 =0$]. This integration constant  is naturally interpreted as the 
rest mass of the particle. 

Substituting (\ref{P1}) back to (\ref{angcon1}), one can rewrite the equation 
of motion of spin in the form:
\begin{equation}
\dot{S}_{\alpha\beta}- U_\alpha U^\gamma\dot{S}_{\gamma\beta} - 
U_\beta U^\gamma\dot{S}_{\alpha\gamma}=0.\label{angcon2}
\end{equation}

\section{New vector variable}

Let us contract (\ref{angcon1}) with $P^\beta$. The result reads:
\begin{equation}
\dot{S}_{\alpha\beta}P^\beta=-U_\alpha P^2 + m_0 P_\alpha.\label{dotS}
\end{equation}
Assuming that the square of the 4-momentum, $P^2:=P_\alpha P^\alpha$, 
is non-zero, it is convenient to introduce a new variable
\begin{equation}
Q_\alpha:={\frac 1{P^2}}P^\beta S_{\alpha\beta},\label{Q1}
\end{equation}

{}From the normalisation condition (\ref{norm}) for the velocity, we find
\begin{equation}
1=U^{\alpha} U_{\alpha} =\left(\frac{m{}_{0} }{P^{2}} \right)^{2} P^{\alpha } 
P_{\alpha } +\dot{Q}^{\alpha } \dot{Q}_{\alpha }.\label{UQ1}
\end{equation}

The equation (\ref{dotS}) takes the form
\begin{equation}
U_\alpha + \dot{Q}_\alpha = {\frac {m_0 P_\alpha}{P^2}}.\label{UQ}
\end{equation}
Recalling (\ref{velo}), we can easily integrate the last equation to find:
\begin{equation}
X_\alpha + Q_\alpha = {\frac {m_0 P_\alpha}{P^2}}\,\tau + 
{\stackrel{(0)} X}_\alpha + {\stackrel{(0)} Q}_\alpha .\label{X1}
\end{equation}

{}From the condition (\ref{cond}) it follows that the new variable
is orthogonal to the 4-velocity,
\begin{equation}
Q_\alpha U^\alpha =0.\label{UQort}
\end{equation}
Hence the vector $Q_\alpha$ is spacelike. On the other hand, by 
construction, $Q_\alpha$ is orthogonal also to the 4-momentum,
\begin{equation} 
Q_\alpha P^\alpha =0.\label{PQort}
\end{equation}
Finally, contracting (\ref{UQ}) with $Q^\alpha$ and using the
condition (\ref{cond}) together with (\ref{PQort}), we obtain
\begin{equation}
Q_\alpha\dot{Q}^\alpha =0.\label{QQ}
\end{equation}

Substituting (\ref{X1}) into (\ref{angcon}) we find a new representation
of the spin tensor: 
\begin{equation}
S_{\alpha\beta}=\mu_{\alpha\beta} + Q_\alpha P_\beta - 
Q_\beta P_\alpha,\label{S1}
\end{equation}
where
\begin{equation}
\mu_{\alpha\beta}=J_{\alpha\beta} - ({\stackrel{(0)} X}_\alpha + 
{\stackrel{(0)}Q}_\alpha)P_\beta + ({\stackrel{(0)} X}_\beta + 
{\stackrel{(0)}Q}_\beta)P_\alpha.\label{mu1}
\end{equation}
is a constant tensor. Comparing with (\ref{Jab}), one can write its
components explicitly in terms of the initial values of spin and momentum:
\begin{equation}
\mu_{\alpha\beta}={\stackrel{(0)}S}_{\alpha\beta} - 
{\frac 1{P^2}}P_\alpha P^\gamma{\stackrel{(0)} S}_{\gamma\beta}
- {\frac 1{P^2}}P_\beta P^\gamma{\stackrel{(0)} S}_{\alpha\gamma}.
\label{mu2}
\end{equation}
In other words, $\mu_{\alpha\beta}$ describes the projection of spin on the
total momentum $P_\alpha$ at the initial moment. If, initially, spin is parallel
to the momentum (i.e., the orthogonal projection vanishes, $P^\gamma{\stackrel
{(0)} S}_{\alpha\gamma} = 0$), $\mu_{\alpha\beta}$ reduces to the tensor of spin. 

Thus, as soon as the time dependence of the variable $Q_\alpha$ is known,
the dynamics of the particle (its position and spin as the functions of
the proper time) is completely described by (\ref{X1}) and (\ref{S1}). It
is worthwhile to stress that $Q_\alpha$ is a composite variable and hence its
initial values are not arbitrary, they are fixed by the initial values of 
momentum and spin.

\section{Physical solutions}

Now, let us find $Q_\alpha(\tau)$ explicitly. {}From (\ref{mu2}) we discover 
another orthogonality relation
\begin{equation}
\mu_{\alpha\beta}P^\beta=0\label{muPort}
\end{equation}
which is crucial to obtain the evolution equation for $Q_\alpha$ alone.
Indeed, taking $U_\alpha$ from (\ref{UQ}), substituting it into the 
condition (\ref{cond}), and making use of the definition (\ref{Q1}) 
and of the relations (\ref{S1}) and (\ref{muPort}), one finds: 

\begin{equation}  
\mu _{\alpha \beta }  \dot{Q}^{\beta } =\left(\frac{\hat{\beta }}{\hat{\alpha }} 
P^{2} +m_{0} \right)Q_{\alpha }.\label{6_24}  
\end{equation}
In the course of this derivation we also used (\ref{PQort}) and (\ref{QQ}). 
We assume that $\hat{\alpha}\neq 0$, the case of the vanishing $\hat{\alpha}$
will be analysed separately. 

Denote the effective mass:

\begin{equation}  
\tilde{m}_{0}:=\frac{\hat{\beta }}{\hat{\alpha }} P^{2} +m_{0}.\label{6_25}  
\end{equation} 
Then, the equation (\ref{6_24}) can be re-written as:

\begin{equation}
\mu_{\alpha\beta}\dot{Q}^\beta - \tilde{m}_{0} Q_\alpha =0.\label{Qosc}
\end{equation}

The system of linear differential equations (\ref{Qosc}) is straightforwardly
integrated. The standard ansatz $Q_\alpha = q_\alpha\,e^{-\,i\omega\tau}$ leads
to the algebraic system for the amplitudes $q_\alpha$ which has nontrivial
solutions when the characteristic determinant is zero,
\begin{equation}
\det\left|\tilde{m}_{0}g_{\alpha\beta} + i\omega\,\mu_{\alpha\beta}\right| = 0.
\end{equation}
The computation of the determinant yields 
\begin{equation}\label{det}
\tilde{m}_{0}^2 - \omega^2\,{\frac 1 2}\,\mu_{\alpha\beta}\mu^{\alpha\beta} = 0,
\end{equation}
because $\varepsilon^{\alpha\beta\mu\nu}\mu_{\alpha\beta}\mu_{\mu\nu}=0$
in view of (\ref{muPort}). 

The nontrivial solutions arise when $\mu_{\alpha\beta}\mu^{\alpha\beta}\neq 0$.
In particular, for $\mu_{\alpha\beta}\mu^{\alpha\beta} > 0$ the general 
solution of (\ref{Qosc}) is an oscillatory motion 
\begin{equation}
Q_\alpha(\tau)= \nu_\alpha\,\cos(\omega\tau) +
\lambda_\alpha\,\sin(\omega\tau),\label{Qsol}
\end{equation}
with the frequency 
\begin{equation}\label{om}
\omega = {\frac {\tilde{m}_{0}}{\sqrt{{\frac 12}
\mu_{\alpha\beta}\mu^{\alpha\beta}}}}.
\end{equation}
Here the constant coefficients $\nu_\alpha, \lambda_\alpha$ are determined
by the initial values of the variables. Explicitly, we find
\begin{equation}
\nu_\alpha = {\stackrel{(0)} Q}_\alpha ={\frac 1{P^2}}\,P^\beta
{\stackrel{(0)} S}_{\alpha\beta},\label{nu0}
\end{equation}
and $\lambda_\alpha$ is constructed below in Sec.~\ref{oscsol}. 
If $\hat{\alpha }$ and $\hat{\beta }$ are such that $\hat{\alpha }U^{\beta } 
+\hat{\beta }P^{\beta }=0$, i.e. $\tilde{m_{0}}$=0, the frequency of vibrations 
$\omega$ is zero, so that the corresponding trajectory is a straight line.   
Also, when $\mu_{\alpha\beta}=0$, the equation (\ref{Qosc}) yields $Q_\alpha=0$, 
and as it follows from (\ref{X1}), the particle moves along the straight line:
\begin{equation}
X_\alpha = {\frac {\tilde{m}_{0}P_\alpha}{P^2}}\,\tau + 
{\stackrel{(0)} X}_\alpha.\label{X2}
\end{equation}
In particular, this also means that $\nu_\alpha=0$, and
$\mu_{\alpha\beta}={\stackrel{(0)}S}_{\alpha\beta}=0$, i.e. it is a spinless case.

Let us assume that $\mu_{\alpha\beta}\neq 0$. Then (\ref{muPort}) is
equivalent to the existence of a 4-vector 
\begin{equation}
\mu^\alpha:={\frac 12}\varepsilon^{\alpha\beta\rho\sigma}
P_\beta\mu_{\rho\sigma},\label{mua}
\end{equation}
which is orthogonal to the 4-momentum,
\begin{equation}
\mu^\alpha P_\alpha =0.\label{muaPort}
\end{equation}
The inverse of (\ref{mua}) reads
\begin{equation}\label{muab}
\mu_{\alpha\beta}={\frac 1{P^2}}\varepsilon_{\alpha\beta\mu\nu}P^\mu\mu^\nu.
\end{equation}
Substituting this into (\ref{Qosc}), one finds
\begin{equation}
{\frac 1{P^2}}\varepsilon_{\alpha\beta\mu\nu}P^\mu\mu^\nu\dot{Q}^\beta - 
\tilde{m}_{0}Q_\alpha =0,\label{Qosc1}
\end{equation}
which immediately yields a new orthogonality relation,
\begin{equation}
\mu^\alpha Q_\alpha =0.\label{muaQort}
\end{equation}

\section{Properties of oscillatory solutions}\label{oscsol}

Summarising the above results, we observe that the vectors
$\{P^\alpha, \mu^\alpha, Q^\alpha\}$ comprise a triplet of the  mutually 
orthogonal vectors at any moment of proper time $\tau$. More precisely, the 
pair of {\it constant} (and {\it mutually orthogonal}, see (\ref{muaPort}))
vectors $P^\alpha, \mu^\alpha$ define a 2-dimensional plane. In view of 
(\ref{PQort}) and (\ref{muaQort}), the variable $Q^\alpha$ always lies in 
the spacelike 2-dimensional plane which is orthogonal to the first plane.
 
We can finalise the integration of (\ref{Qosc}) as follows. Inserting 
(\ref{Qsol}) in (\ref{Qosc1}), one finds the two relations between the
four constant vectors:
\begin{eqnarray}
\lambda_\alpha &=& -\,{\frac \omega {\tilde{m}_{0}P^2}}
\,\varepsilon_{\alpha\beta\mu\nu}\,\nu{}^\beta P^\mu\mu^\nu,\label{Q01}\\
\nu_\alpha &=& {\frac \omega {\tilde{m}_{0}P^2}}\,\varepsilon_{\alpha\beta\mu\nu}
\,\lambda^\beta P^\mu\mu^\nu.\label{Q0}
\end{eqnarray}
This shows that the constant vectors $\nu_\alpha$ and
$\lambda_\alpha$ are orthogonal to each other, 
\begin{equation}
\nu_\alpha\lambda^\alpha=0,\label{QQort}
\end{equation}
and to $P^\alpha, \mu^\alpha$ as well,
\begin{equation}
\nu_\alpha P^\alpha = \nu_\alpha\mu^\alpha =
\lambda_\alpha P^\alpha = \lambda_\alpha\mu^\alpha =0.\label{QQPort}
\end{equation}
Substituting (\ref{Q01}) into (\ref{Q0}), we find:
\begin{equation}
\nu_\alpha = - \left({\frac {\omega}{\tilde{m}_{0}P^2}}\right)^2
\varepsilon_{\alpha\beta\mu\nu}\varepsilon^{\beta\gamma\rho\sigma}
\nu_\gamma P_\rho\mu_\sigma P^\mu\mu^\nu =
- \left({\frac {\omega}{\tilde{m}_{0}}}\right)^2{\frac {\mu_\beta\mu^\beta}{P^2}}
\nu_\alpha, \label{om1}
\end{equation}
where we repeatedly used the orthogonality relations (\ref{QQPort}). 
As one can straightforwardly verify from (\ref{mua}) and (\ref{muab}), 
\begin{equation}
{\frac 1 2}\mu_{\alpha\beta}\mu^{\alpha\beta}=
-\,{\frac {\mu_\alpha\mu^\alpha} {P^2}},\label{mumu}
\end{equation}
and hence (\ref{om1}) is identically satisfied when the frequency
$\omega$ is given by (\ref{om}).

Hence, provided the initial conditions for the momentum 
${\stackrel{(0)}P}_\alpha$ and spin ${\stackrel{(0)}S}_{\alpha\beta}$ are 
given, one can construct a constant spacetime frame represented by the 
four vectors:
\begin{equation}
P^\alpha, \quad\quad \mu^\alpha, \quad\quad \nu{}^\alpha, 
\quad\quad \lambda^\alpha,\label{consframe}
\end{equation}
as shown in (\ref{mua}), (\ref{Q1}) and (\ref{Q01}). The four vectors 
(\ref{consframe}) are all orthogonal to each other; $P^\alpha$ is timelike
(i.e., $P^2=P_\alpha P^\alpha > 0$), whereas the three other are spacelike 
(thus having negative length square). One can show that the lengths of the
three spacelike vectors are proportional to each other. Indeed, squaring
(\ref{Q01}), we find 
\begin{equation}
\lambda_\alpha\lambda^\alpha = \nu_\alpha\nu{}^\alpha.\label{n2l2}
\end{equation}
Then from (\ref{Qsol}) we immediately obtain:
\begin{equation}
\dot{Q}_\alpha\dot{Q}^\alpha=\omega^2 Q_\alpha Q^\alpha,
\end{equation}
and then using (\ref{UQ1}), 
\begin{equation}
1=U_\alpha U^\alpha = {\frac {m_0^2}{P^2}} + \omega^2 Q_\alpha Q^\alpha.
\end{equation}
{}From this we find finally (insert (\ref{om}) and (\ref{mumu})):
\begin{equation}\label{QQmu}
\nu_\alpha\nu^\alpha =\frac{\mu ^{\alpha } \mu _{\alpha } }{P^{2} } 
\left(\frac{m_{0} ^{2} }{P^{2} \left(\frac{\hat{\beta }}{\hat{\alpha }} 
P^{2} +m_{0} \right)^2} -\frac{1}{\left(\frac{\hat{\beta }}{\hat{\alpha }} 
P^{2} +m_{0} \right)^2} \right)
\end{equation}
This shows, since $\nu_\alpha$ is spacelike, that one must have
\begin{equation}
P^2 < m_0^2.
\end{equation}

The dynamics of spin is given by (\ref{S1}). In order to get a further insight
(motivated by the quantum mechanical nature of spin, \cite{ost}), one usually 
introduces the 4-vector of spin by $S^\alpha = {\frac 12}\varepsilon^{\alpha\beta
\rho\sigma}U_\beta S_{\rho\sigma}$. Then by using the Mathisson-Papapetrou 
equations of motion and the results obtained above, we straightforwardly 
verify that the spin vector precesses as
\begin{equation}
\dot{S}^\alpha = \Omega^\alpha{}_\beta\,S^\beta.\label{precess}
\end{equation}
The precession angular velocity is $\Omega_{\alpha\beta} = -\omega\hat{\mu}
{}_{\alpha\beta}$, where $\hat{\mu}{}_{\alpha\beta} = \mu_{\alpha\beta}/\sqrt{
{\frac 12}\mu_{\rho\sigma}\mu^{\rho\sigma}}$ is the unit tensor giving the 
direction of the precession axis.

\section{Special cases}

Let us summarise the results on the dynamics of a free particle with spin. 
There are several qualitatively different cases which arise for the different
limits of the supplementary condition and for the different initial conditions.

\subsection{Tulczyjew condition}

When $\hat{\alpha}=0$ the condition (\ref{cond}) becomes the Tulczyjew's 
condition:
\begin{equation}
S_{\alpha\beta} P^\beta =0
\end{equation}
Consequently, the $Q_\alpha$ as defined in the (\ref{Q1}) vanishes at any 
$\tau$, meaning that the trajectories are straight lines described by 
\begin{equation}
X^\alpha ={\frac{P^\alpha}{P^2}}\tau + {\stackrel{(0)}X}{}^\alpha,
\end{equation}
together with constant momentum and spin
\begin{equation}
P_\alpha = {\stackrel{(0)}P}{}_\alpha,\qquad
S_{\alpha\beta} = {\stackrel{(0)}S}{}_{\alpha\beta}.
\end{equation}

\subsection{Frenkel condition}

When $\hat{\beta }$=0, the Frenkel condition,is recovered:
\begin{equation} 
S_{\alpha\beta}U^\beta = 0.\label{Frenkel} 
\end{equation}
Under the Frenkel condition, $m_0$ is constant along the world line and 
the oscillation frequency is 
\begin{equation}
\omega = {\frac {m_{0}}{\sqrt{{\frac 12}\mu_{\alpha\beta}
\mu^{\alpha\beta}}}}.\label{omtfrenkel}
\end{equation}

One can establish several other useful relations between spin and momentum.
For example, directly from (\ref{S1}) one sees:
\begin{equation}
{\frac 12}S_{\alpha\beta}S^{\alpha\beta}={\frac 12}\mu_{\alpha\beta}
\mu^{\alpha\beta} + P^2 Q_\alpha Q^\alpha,
\end{equation}
and thus using (\ref{QQmu}), we find
\begin{equation}
{\frac 12}S_{\alpha\beta}S^{\alpha\beta}={\frac 12}\mu_{\alpha\beta}
\mu^{\alpha\beta}{\frac {P^2}{m_0^2}}.\label{mu2p2}
\end{equation}
This suggests another form for the frequency:
\begin{equation}
\omega={\frac {|P|}{\sqrt{{\frac 1 2}S_{\alpha\beta}S^{\alpha\beta}}}},\label{om2}
\end{equation}
where $|P|=\sqrt{P_\alpha P^\alpha}$. Interesting to note that one can
rewrite the above equation in the form
\begin{equation}
|P|=|S|\omega,
\end{equation}
which with $|S|=\sqrt{{\frac 12}S_{\alpha\beta}S^{\alpha\beta}}$ resembles
the well-known wave-particle relation ($E=h\nu$) if one estimates 
$|S|=\hbar/2$ and replaces $\omega=2\pi\nu$. With these identifications,
$\omega$ coincides with the Zitterbewegung frequency derived in the quantum
theory.

\subsection{Straight line motion}

We have seen that the cases of spinless particles, as well as the conditions 
$\hat{\alpha}=0$ and $\hat{\alpha }U^{\beta } +\hat{\beta }P^{\beta }=0$ 
correspond to the straight line motion. 

Furthermore, one can show that the particle with spin moves monotonously along 
a straight line if and only if, at an initial moment $\tau=0$
\begin{equation}
{\stackrel{(0)}S}_{\alpha\beta}{\stackrel{(0)}P}{}^\beta =0.
\label{case1}
\end{equation}

Then the particle's position is described by (\ref{X2}), while the spin
is constant,
\begin{equation}
S_{\alpha\beta}=\mu_{\alpha\beta}.\label{S2}
\end{equation}
The proof is as follows. The initial condition (\ref{case1}) is equivalent 
to $\nu_\alpha=0$, so when this is satisfied, we find from (\ref{Qsol}), 
(\ref{Q01}) that $Q_\alpha=0$ for any $\tau$. Hence, (\ref{X1}) reduces to 
(\ref{X2}) and (\ref{S1}) to (\ref{S2}). On the other hand, suppose the 
particle moves along a straight line $X_\alpha = C_\alpha\tau + {\stackrel{(0)} 
X}_\alpha$ with some constant vector $C_\alpha$. Then $U_\alpha=C_\alpha$, 
and thus $U^\beta\dot{S}_{\alpha\beta} = {\frac d{d\tau}}(U^\beta S_{\alpha\beta})
=0$. Hence the equation (\ref{P1}) reduces to $P_\alpha=m_0U_\alpha$ which 
gives the constant vector $C_\alpha = P_\alpha/m_0$, and we again recover 
(\ref{X2}). At the same time, the Frenkel condition (\ref{cond}) then 
yields (\ref{case1}).

\subsection{Classical Zitterbewegung}

When, at an initial moment $\tau=0$,
\begin{equation}
{\stackrel{(0)}S}_{\alpha\beta}{\stackrel{(0)}P}{}^\beta\neq 0,
\label{case2}
\end{equation}
the particle's motion is described by (\ref{X1}) and (\ref{S1}), where
the variable $Q_\alpha$ oscillates according to (\ref{Qsol}) with the
frequency (\ref{om}). This picture is a classical counterpart of the
{\it Zitterbewegung} of a Dirac particle in relativistic quantum mechanics.
There is a natural global orthogonal frame (\ref{consframe}) in which this 
motion is described in a most clear way. 

It seems worthwhile to note that the explicit 4-dimensional picture makes
more transparent the details of the classical oscillatory motion. In 
particular, contrary to the statements which can be encountered in the
3-dimensional analyses (see, e.g., \cite{wey1,cor}), the dynamics of spinning
particle is not a mere superposition of a motion along a line with a rotation
around the direction of momentum or around the direction of spin. More 
precisely, we have learned that the particle's rotary motion is confined 
to the plane $(\nu_\alpha, \lambda_\alpha)$ orthogonal to {\it both} the
momentum and spin.

\section{Non-physical solutions}

One can verify that the relation (\ref{mu2p2}) holds true {\it in general}, 
and not only for the above case when $\mu_{\alpha\beta}\mu^{\alpha\beta} > 0$ 
(then this relation confirms that the momentum in timelike). 

Correspondingly, we read from (\ref{mu2p2}) that assuming the unphysical 
{\it spacelike} momentum, i.e., $P^2 < 0$, we automatically find that 
$\mu_{\alpha\beta}\mu^{\alpha\beta} < 0$. Returning now to the linear
system (\ref{Qosc}), we see that instead of the oscillatory solutions
(\ref{Qsol}) we find the hyperbolic motion:
\begin{equation}
Q_\alpha(\tau)= \nu_\alpha\,\cosh(\overline{\omega}\tau) +
\lambda_\alpha\,\sinh(\overline{\omega}\tau),\label{Qsol1}
\end{equation}
with the ``acceleration'' parameter
\begin{equation}
\overline{\omega} = {\frac {\tilde{m}_{0}}{\sqrt{-\,{\frac 12}\mu_{\alpha\beta}
\mu^{\alpha\beta}}}}.\label{omt}
\end{equation}
One can immediately see that the constant vectors $\nu_\alpha$ and 
$\lambda_\alpha$ satisfy the relations
\begin{equation}
\nu_\alpha\,\nu^\alpha = -\,\lambda_\alpha\,\lambda^\alpha,\qquad
\nu_\alpha\,\lambda^\alpha = 0.
\end{equation}
Consequently, since the variable $Q^\alpha$ is always spacelike, we
discover that the vector $\nu_\alpha$ is also spacelike, whereas 
$\lambda_\alpha$ is timelike. Recalling that (\ref{mumu}) in all cases
describes the vector $\mu_\alpha$ as spacelike, we thus again end with
the global vierbein (\ref{consframe}), where however the role of the 
timelike leg is played by the vector $\lambda_\alpha$ instead of the
momentum. 

The existence of such a solution was mentioned recently 
in \cite{kh}. It is clear that the self-accelerated solutions of the
type (\ref{Qsol1}) (recall that the particle is not affected by any
external forces) are non-physical and one should exclude them. This
imposes a transparent constraint on the possible values of the momentum:
$P^2 > 0$.

\section{Conclusion}

In this paper we have presented a complete explicitly covariant 4-dimensional 
description of the dynamics of a free classical particle with spin in
the framework of the special relativity theory. The crucial step in
the analysis of the problem was provided by the introduction of the
new vector variable $Q^\alpha$. {}From (\ref{S1}) and (\ref{X1}), one
can notice a certain analogy with the definition of the mean spin and 
mean position operators in relativistic quantum mechanics \cite{mean}. Finally,
we demonstrated that, contrary to the earlier statements \cite{tulczyjew,7r},
the dynamics of a spinning particle (both the motion of spin and the 
trajectory) under the Frenkel condition is uniquely determined by the 
initial conditions, i.e., by the values of ${\stackrel{(0)}X}{}^\alpha, 
{\stackrel{(0)}P}{}_\alpha, {\stackrel{(0)}S}_{\alpha\beta}$. This important
property is valid also for the curved spacetimes, as we will show elsewhere.

\end{document}